\documentclass[12pt,preprint]{emulateapj}
\usepackage{amsmath}
\usepackage{framed}

\newcommand{\beq}{\begin{equation}}
\newcommand{\eeq}{\end{equation}}

\newcommand{\ms}{M_*}
\newcommand{\msol}{{\rm M_{\odot}}}

\def\dim#1{\mbox{\,#1}}

\def\H2{{{\rm H}_2}}
\def\HI{{\rm H\,I}}

\def\Msun{{\rm M}_{\odot}}
\def\Zsun{{\rm Z}_{\odot}}

\def\Sntr{\Sigma_{\HI+\H2}}

\def\yeff{y_{\rm eff}}

\def\hide#1{}
\def\dim#1{\mbox{\,#1}}

\begin{document}

\title{Ultra-faint dwarf galaxies as a test of early enrichment\\ and metallicity-dependent star formation }

\author{Konstantinos Tassis\altaffilmark{1}, Nickolay Y. Gnedin\altaffilmark{2,3,4} and Andrey V. Kravtsov\altaffilmark{3,4,5}}
\altaffiltext{1}{Jet Propulsion Laboratory, California Institute of Technology, Pasadena, CA 91109, USA}
\altaffiltext{2}{Particle Astrophysics Center, Fermi National Accelerator Laboratory, Batavia, IL 60510, USA}
\altaffiltext{3}{Kavli Institute for Cosmological Physics and Enrico Fermi Institute, The University of Chicago, Chicago, IL 60637, USA}
\altaffiltext{4}{Department of Astronomy \& Astrophysics, The University of Chicago, Chicago, IL 60637 USA}
\altaffiltext{5}{Enrico Fermi Institute, The University of Chicago, Chicago, IL 60637, USA}
\begin{abstract}
The tight relation of star formation with molecular gas indicated by observations and assumed in recent models implies that the efficiency with which galaxies convert their gas into stars depends on gas metallicity. This is because abundance of molecular hydrogen is sensitive to abundance of dust, which catalyses formation of H$_2$ and helps to shield it from dissociating radiation. In this study we point out that in the absence of significant pre-enrichment by Population III stars forming out of zero metallicity gas, such H$_2$-based star formation is expected to leave an imprint in the form of bi-modality in the metallicity distribution among dwarf galaxies and in the metallicity distribution of stars within individual galaxies. The bi-modality arises because when gas metallicity (and dust abundance) is low, formation of molecular gas is inefficient, the gas consumption time scale is long, and star formation and metal enrichment proceed slowly. When metallicity reaches a critical threshold value star formation and enrichment accelerate, which leads to rapid increase in both stellar mass and metallicity of galaxies. We demonstrate this process both using a simple analytical model and  full cosmological simulations. In contrast, observed metallicity distribution of dwarf galaxies or stars within them are not bi-modal. We argue that this discrepancy points to substantial early stochastic pre-enrichment by population III stars to levels $Z\sim 10^{-2}Z_\odot$ in dense, star forming regions of early galaxies.
\end{abstract}



\section{Introduction}
\label{sec:intro}

The observed correlation between stellar mass and metallicity, established for a wide range of galaxy masses, morphologies, and metallicities, is an important diagnostic
in the investigation of physical processes that shape the baryonic component of galaxies. 
\cite{DW03} find a correlation of the form $Z\propto M_{\ast}^{\alpha_Z}$ with $\alpha_Z\approx 0.4$ for the Local Group dwarf spheroidal galaxies. Analysis of the SDSS data by
\citet{Trem04} (see also \citealp{gallazzi_etal05}) indicates that the
slope of the scaling increases with decreasing $\ms$: it is quite flat
for $\ms \gtrsim 3 \times 10^{10} \msol$, while it steepens for lower
masses and exceeds the \citet{DW03} value by $M\sim {\rm few}\times
10^8 \msol$. \citet{lee06}, on the other hand, find that at even lower stellar
masses the scaling becomes less steep, with a slope of $\alpha_Z=0.29\pm 0.03$
for the dwarf galaxies in their sample ($10^6 {\rm \, \msol } \lesssim
M_* \lesssim$ few $\times 10^8 {\rm \, \msol}$). Most recently,
\citet{dsh:klsc11} find that at even lower stellar masses ($M_{\ast}\lesssim 10^5\rm\,\msol$) the relation becomes flat, $\alpha_Z\approx 0$.

The stellar mass-metallicity relation of low-metallicity systems is sensitive to details of star formation, enrichment, and metal transport  during the earliest stages of galaxy evolution.  For example, it was argued that $M_{\ast}-Z$ is sensitive to blow-out of metals via galactic winds \citep[e.g.,][]{Tetal03,DW03,Kob07} or to increasing inefficiency of star formation with decreasing galaxy mass
\citep{brooks_etal07,derossi_etal07,tkg08, gk10}. The latter can be due to star formation regulation by feedback \citep[e.g.,][]{brooks_etal07} or by processes controlling formation of dense, star forming regions within ISM \citep{tkg08}. One such key process is regulation of molecular gas formation by dust abundance. 
Both observations \citep{misc:wb02,ism:lwbb08,sfr:blwb08,sfr:gtgs10,sfr:blwb11} and theoretical considerations \citep[e.g.,][]{sfr:klm11} suggest that star formation is tightly correlated with the mass of gas in molecular phase.  The latter is sensitive to dust abundance,  which catalyses the formation of H$_2$ and helps to shield it from dissociating radiation \citep[e.g.,][]{spaansN97,hf02,sfr:kmt09a,ng:gtk09,ng:gk11,ays11}. 

One of the implication of this tight relation is a ``catch-22'' situation that exists during early stages of galaxy evolution and chemical enrichment: stars are needed to produce metals and dust, while dust is needed to facilitate star formation. Star formation can thus be expected to proceed first in the central regions of galaxies, where gas surface density is high and it is easier for H$_2$ to form even with small amount of dust. As central regions are enriched, the metals can be mixed with gas at larger radii either by internal mixing processes or via mergers of gaseous disks. As lower surface density regions are enriched, formation of molecular gas and star formation will become more efficient at larger radii within galaxies. Given that this sequence requires a certain amount of time, which can be prolonged by loss of metals via feedback-driven winds, star formation in galaxies can be expected to be delayed or even completely suppressed. The delay or suppression depend on the surface density of gas within galaxies and can thus be  expected to be the largest in dwarf galaxies, which have the smallest gas surface densities. Simulations \citep{gk10,kuhlen_etal11} and analytical models \citet{gals:kd11} show that strong suppression of star formation via such processes occurs in halos of total mass $M_{\rm h}\lesssim 10^{10}\rm\, \msol$.

In this paper we examine how such processes affect the metallicities of galaxies and the metallicity distribution of stars within them in more detail
by using both simple analytical model (\S~\ref{sec:theory}) and full
cosmological simulations (\S~\ref{sims}) based on the model for H$_2$ formation and H$_2$-based star formation introduced in \citet{ng:gtk09} and \citet{ng:gk11}.
We show that when star formation is tied to molecular gas,  bi-modality in metallicity arises generically due to the existence of a characteristic gas surface density below which star formation efficiency is suppressed. However, as we discuss in \S~\ref{comparison}, observations do not show evidence for such bi-modality in high redshift or local low-metallicity systems. We interpret this apparent discrepancy between theoretical expectation and observations as indication that gas within regions where first Population II stars were formed was pre-enriched by Population III stars to metallicities as high as $\sim 10^{-2}\, Z_{\odot}$. Such interpretation is consistent with models of dwarf Milky Way satellites, the metallicity distribution of the Milky Way halo stars, and observations of metallicity distribution of the damped Ly$\alpha$ absorption systems, which also indicate that pre-enrichment at least to $Z\sim 10^{-3}\, Z_{\odot}$ is required \citep{dsh:hitb06,roll09,dsh:klsc11,kbbh11}. Furthermore, abundance
patterns in very low metallicity systems are in good agreement with
those expected from the Population III stars \citep{bl03,simon10leo,cooke11}. Significant pre-enrichment is also suggested by simulations of feedback by the Population III stars \citep{gals:wtna11}.

\section{Metallicity-mediated star formation and enrichment history of galaxies}
\label{sec:theory}

We begin by illustrating how bi-modal metallicity distribution can be expected to arise generically both for the distribution of metallicities within galaxies and the distribution of average metallicities of galaxies. We do so using a
simple analytical model which, nevertheless, captures the relevant physics in a qualitative way.

Studying the environmental effects of $\H2$-based star formation
models, \cite{gk10} found that the surface density of the star
formation rate as a function of the total neutral gas surface density,
$\Sntr$, can be well approximated as
\begin{equation}
  \Sigma_{\rm SFR} = - \dot{\Sigma}_{\HI+\H2} = \frac{\Sntr}{\tau_{\rm SF}}
  \left(1+\frac{\Sigma_{\rm crit}}{\Sntr}\right)^{-2}.
  \label{eq:ksfit}
\end{equation}
where $\dot{\Sigma}_{\HI+\H2}$ is the neutral gas
consumption rate per unit area and $\Sigma_{\rm crit}$ is a characteristic ``threshold''
surface density of neutral gas below which the $\Sigma_{\rm SFR}-\Sigma_{\HI+\H2}$ relation steepens as
 gas becomes predominantly atomic at lower surface densities.
Defining $\Sigma_t=\Sigma_{\HI+\H2}+\Sigma_{\ast}$, where $\Sigma_{\ast}$ is stellar surface density, the (neutral) gas fraction at each location is $ f_g \equiv
\Sntr/\Sigma_t$, and the evolution of $f_g$ at constant $\Sigma_t$ is
described by the following differential equation:
\begin{equation}
  \frac{df_g}{dt} = -\frac{f_g}{\tau_{\rm SF}}
  \left(1+\frac{\Sigma_{\rm crit}}{f_g\Sigma_t}\right)^{-2}.
  \label{eq:fgevol}
\end{equation}
In general, the threshold surface density $\Sigma_{\rm crit}$ is a complex
function of the dust-to-gas ratio (which we assume to be linearly
proportional to the gas metallicity $Z$ in this paper), and the
interstellar radiation field. However, for low $Z$, $\Sigma_{\rm crit}$ is
approximately inversely proportional to $Z$, with the coefficient of
proportionality being somewhat dependent on the value of the
interstellar radiation field [see Figure 6 of \citet{gk10}]. Radiation
fields in high-redshift galaxies are generally higher than in local
galaxies \citep{cppp09}, so in the following we assume $\Sigma_{\rm crit}
\approx 10\,\Msun\dim{pc}^{-2}/Z$, which is appropriate for radiation
fields 10 to 100 times higher than in the Milky Way. Since
$\Sigma_{\rm crit}$ depends only on approximately the square root of the 
radiation field [see eq. 14 of \citet{gk10}], the range above corresponds 
to a factor of 3 uncertainty in $\Sigma_{\rm crit}$.

The gas fraction and the gas metallicity are related through
\[
  Z = Z_0 + \yeff \mbox{ln}(1/f_g),
\]
where $Z_0$ is the ``primordial'' metallicity (i.e. the metallicity
from star formation that is uncorrelated with the dust abundance),
which may reach values exceeding $10^{-3}$ solar
\citep{gals:wtna11}, and $\yeff$ is the so-called ``effective yield'',
which, in a closed-box approximation, is equal to the true metal
yield. Assuming that the effective yield is a weak function of time
(or, equivalently, gas metallicity), we can derive a single equation
for the gas metallicity as a function of time,
\begin{equation}
  \frac{dZ}{dt} = \frac{\yeff}{\tau_{\rm SF}}
  \left(1+\frac{e^{(Z-Z_0)/\yeff}}{qZ}\right)^{-2},
  \label{eq:zevol}
\end{equation}
where 
$  q \equiv \Sigma_t/(10\,\Msun\dim{pc}^{-2})$.

The solution to Equation (\ref{eq:zevol}) can be obtained analytically for
the limit $Z \ll \yeff$, if we introduce a new time variable
$ dx = (\yeff/\tau_{\rm SF})dt$,
which ``hides'' all complex dependence of $\tau_{\rm SF}$ and $\yeff$ on
other physical quantities. In general, however, $\tau_{\rm SF}$ will
depend on the gas metallicity, so finding a solution of $Z$ as a
function of $x$ only produces an implicit dependence of $Z$ on the
actual physical time $t$. For simple estimates, however, we can assume
that $\tau_{\rm SF} \approx 1.5-2\dim{Gyr}$
\citep{sfr:blwb11,sfr:gtgs10}.

With these simplifying assumptions, Equation (\ref{eq:zevol}) 
reduces to
\begin{equation}
  \frac{dZ}{dx} = 
  \left(1+\frac{1}{qZ}\right)^{-2},
  \label{eq:zevolsim}
\end{equation}
whose implicit solution is
\begin{equation}
  Z - \frac{2}{q}\ln\,Z - \frac{1}{q^2Z} = x + Z_0 - 
  \frac{2}{q}\ln\,Z_0 - \frac{1}{q^2Z_0}.
  \label{eq:zsol}
\end{equation}

Solution (\ref{eq:zsol}) exhibits different behavior for different
values of $Z_0$ and $q$. If $qZ_0 \ll 1$, the metallicity remains just
above the ``primordial'' value $Z_0$ for a long time, $x\gg1$, keeping
galaxies in the low-metallicity, low-stellar-mass part of the
parameter space. However, if $qZ_0 \gg 1$, $Z$ is growing approximately
linearly with time, $Z \approx x+Z_0$, and the gas metallicity
reaches the value $Z=\yeff$ (recall that the solution is only valid
for $Z < \yeff$) in a time interval $ \Delta t = \tau_{\rm SF}$.
Finally, for $Z>\yeff$, the exponential term in Equation
(\ref{eq:zevol}) becomes non-trivial, and the growth rate of $Z$ slows
down considerably and eventually becomes only logarithmic.
\begin{figure}
\begin{center}
\epsscale{0.7}
\plotone{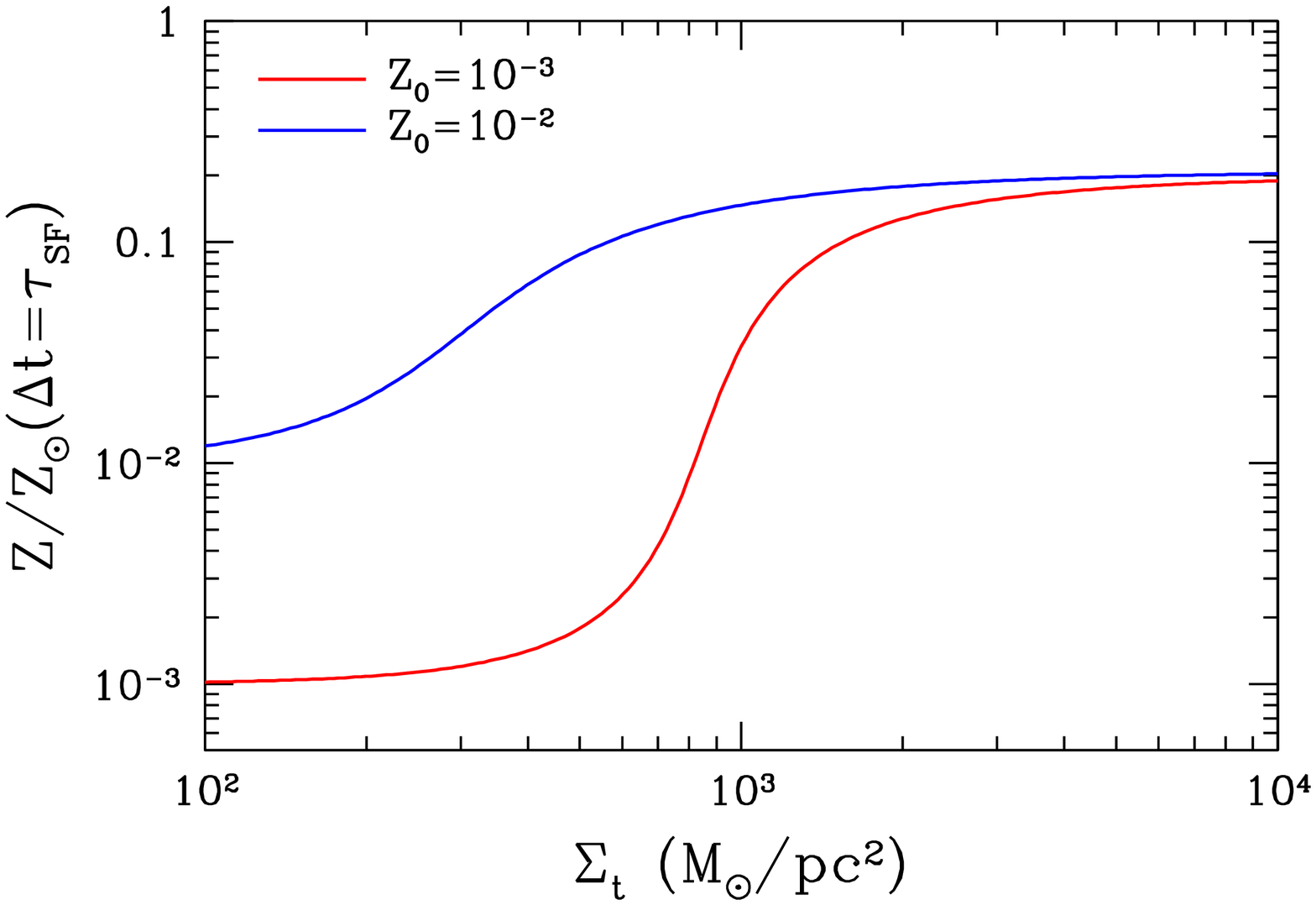}\\
\plotone{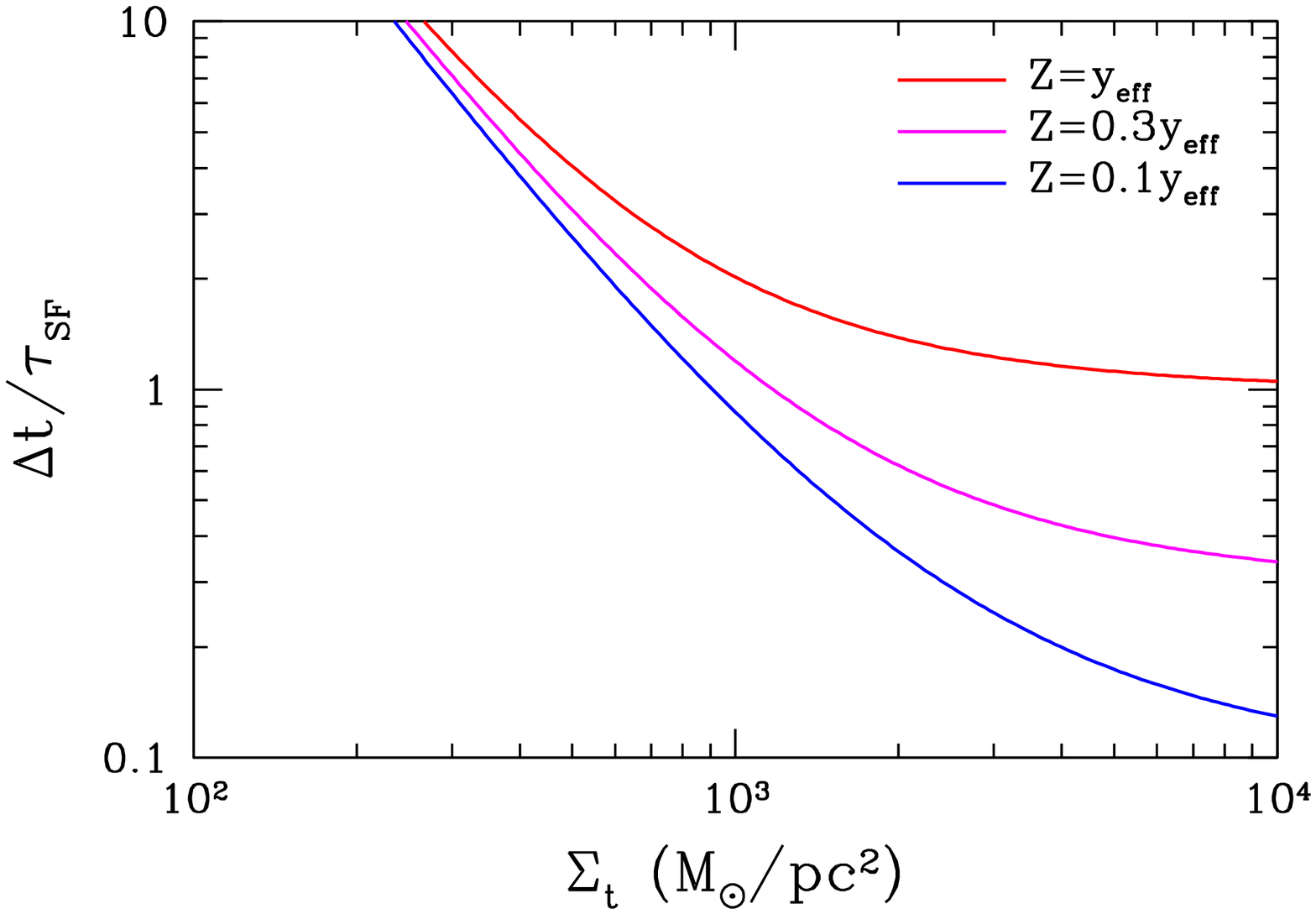}
\end{center}
\caption{\label{fig:ansol} Top panel: metallicity (in solar units) as a
function of the total surface density at $\Delta t = \tau_{\rm SF}$.
Bottom panel: time interval required to reach a particular metallicity as a
function of the total neutral gas surface density. The value of $\yeff=0.2Z_\odot$ is
assumed in the presented model calculations \citep{sfr:d07}.}
\end{figure}
Figure \ref{fig:ansol} illustrates the solution (\ref{eq:zsol}) as a
function of the total surface density for $Z$ at constant $t$ and $t$
at constant $Z$\footnote{Changing $\Sigma_{\rm crit}$ in our solution would translate the curves horizontally by the corresponding factor, but it would not change their shape.}. The transition between the two regimes (low
$\Sigma_t$, low $Z$ and high $\Sigma_t$, high $Z$) is rather rapid when the value of $Z_0$ is low, so
the two regimes are well separated. The transition becomes shallower as $Z_0$ increases. 

It is this transition that is the source of bi-modality in the distribution 
of galaxy metallicities when $Z_0$ is low. Given that surface density of gaseous disks is expected to decrease with decreasing halo mass \citep{mo_etal98}, halos below a certain mass will host disks with surface density smaller than the above transition value. Star formation and metal enrichment would proceed in such disks at a slow rate and they would stagnate in the low $M_{\ast}$, low $Z$ regime. Conversely, gas disks with high surface densities, hosted by sufficiently massive halos, will form substantial mass of stars and will reach high metallicity within period of $\lesssim \tau_{\rm SF}\approx 1.5-2$~Gyrs, at least within their central regions. 
A qualitatively similar effect can be expected for the metallicity distribution of stars within individual galaxies, for systems that spend significant fraction of their evolution in the stagnant, low-metallicity phase. While a galaxy is in this phase, metallicity evolves slowly and stellar mass that is building up will predomninantly consist of low $Z$ stars. If at some point in its evolution either the surface density of its gas disk or the metallicity increase to the level required by the condition above, the galaxy will undergo dramatic increase in star formation rate and will rapidly self-enrich itself producing a population of high $Z$ stars. Given that the transition is rapid, there will be a relative paucity of stars at intermediate metallicities and the metallicity distribution will be bi-modal.   
Note that pronounced bi-modality can be expected only for galaxies that produced significant fraction of stars in both regimes. Clearly, if a galaxy has always evolved in the stagnant regime, it will only have low metallicity stars without a high $Z$ peak, while the low $Z$ peak may be insignificant in a galaxy that formed most of its stars after the transition. 

The simple model and arguments presented above show that bi-modality in metallicity distribution is generically expected in galaxies of a certain mass range.  
We now turn to simulations that include the process of disk formation and H$_2$-based star formation self-consistenly and demonstrate that the expected bi-modality does indeed arise naturally for simulated galaxies.

\begin{figure}
\plotone{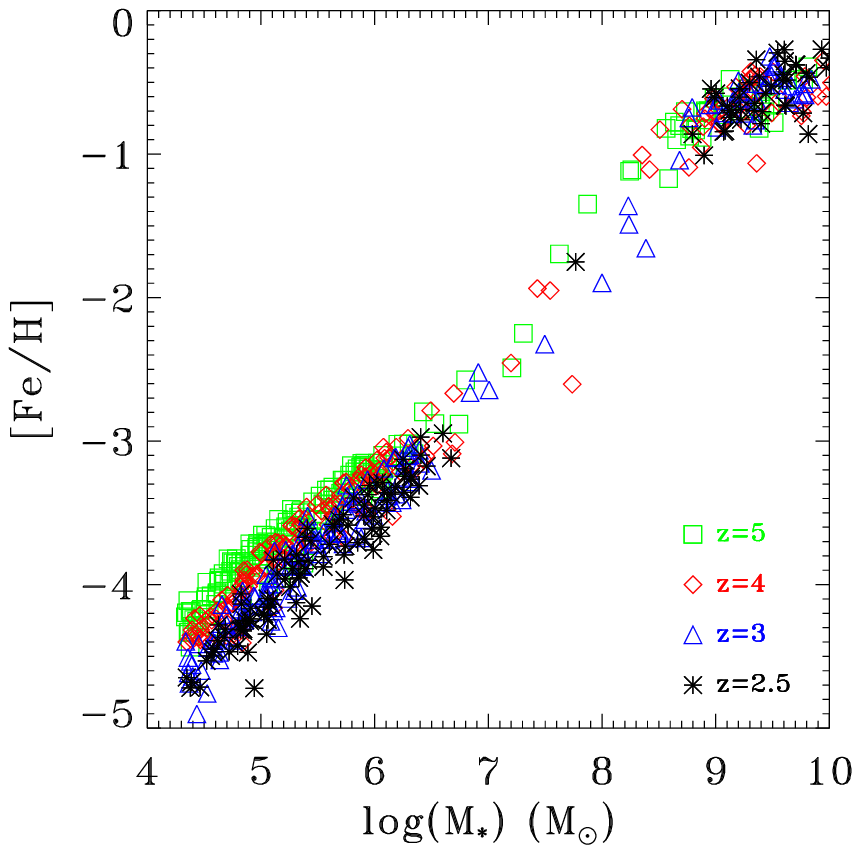}
\caption{\label{fig:mz} Mass-metallicity relation for simulated galaxies at
different redshifts (symbols/colors as in legend).
\hide{Observational data are overplotted in both panels. The blue
crosses are iron metallicity observations in local dwarf galaxies from
\citet{dsh:klsc11}; magenta stars are from \cite{kalirai10}. The
vertical solid line and the left vertical dashed line indicate the
lower mass limit for oxygen abundance observations of \cite{Trem04}
and \cite{lee06}. The latter study includes galaxies with masses up to
the right dashed line.}}
\end{figure}

\section{Metallicity bi-modality in simulations with H$_2$ based star formation}\label{sims}

\begin{figure*}
\plotone{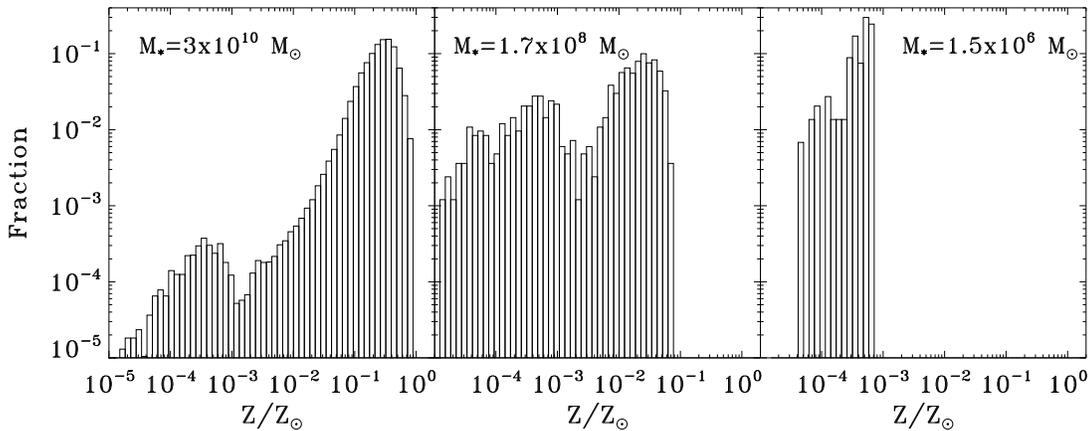}
\caption{\label{Hist} Metallicity distributions of stars in three
simulated galaxies with different stellar masses. 
 The distributions in the larger and medium-mass galaxies (left and middle panels) are 
clearly bimodal.The extremely-low stellar mass galaxy does not have any stars
of metallicity higher than $10^{-3}$ solar. }
\end{figure*}

To examine detailed predictions for the metallicity distribution, we
use a set of high-resolution simulations of WMAP5\footnote{$\Omega_M =
0.28$, $\Omega_B = 0.046$, $h=0.7$, $n_S=0.96$, $\sigma_8=0.82$.}
cosmology \citep{cosmo:kdnb09} carried out with the Adaptive
Refinement Tree (ART) code \citep{misc:k99,misc:kkh02,sims:rzk08}. The
detailed description of the simulation is presented elsewhere
\citep{zemp_etal11}. The simulation volume of $25.6h^{-1}\dim{Mpc}$
includes several regions of interest which are sampled in the initial
conditions with the effective resolution of $2048^3$, resulting in the
dark matter particle mass of $2\times10^5\Msun$. Each region of
interest is allowed to refine dynamically up to the comoving spatial
resolution of $200 h^{-1}\dim{pc}$, which is equivalent to the spatial
resolution in physical units of $(57,70,90,95)\dim{pc}$ at
$z=(4,3,2.5,2)$. Regions of interest are chosen from a lower
resolution dissipationless simulation to include $5R_{\rm vir}$
spheres around six galaxies with $z=0$ masses between $10^{12}$ and
$10^{13}$ solar masses.

The physical processes modelled in the simulation are described in \citet{ng:gk10,ng:gk11}. In particular, we include
gas cooling, including cooling by metals, molecular hydrogen, and
dust, a phenomenological model for molecular hydrogen formation, full
time-dependent and spatially variable 3D radiative transfer of
ionizing and Lyman-Werner band radiation using the Optically Thin
Variable Eddington Tensor (OTVET) approximation of \citet{ga01}, star
formation in the molecular gas using the \citet{sfr:kt07} recipe, and (in
our fiducial run) the supernova feedback modelled as thermal energy
injection over $20\dim{Myr}$ after the formation of a stellar population.

Our simulation does not resolve the formation of the Population III stars
and associated enrichment. However, this initial enrichment and the dust that results from it 
are required for a more widely spread formation of molecular gas and the Population II star formation.
Therefore, we model the enrichment by the unresolved Population III stars by assuming that a gas dense enough to form molecules has a minimum amount of cosmic dust
corresponding to dust-to-gas ratio $10^{-3}$ times smaller than the Milky Way value \citep[i.e., $D_{\rm MW}=10^{-3}$ in the notation of][]{ng:gk11}. Given that we assume that the dust-to-gas ratio scales linearly with gas metallicity, this assumption is equivalent to
setting the initial metallicity to $Z_0=10^{-3}\Zsun$ in the notation of the previous section. This ``dust
floor'' is only accounted for in computing the molecular hydrogen
abundance. It is not added to the actual metal abundance followed in the simulation and is not taken into account in calculation of cooling rates, for example.
However, in the discussion below we will examine results by introducing explicit metallicity floor during post-processing of simulations. 

Star formation in the simulations is based on abundance of molecular gas, as described in \citet{ng:gtk09} and \citet{ng:gk11}. Our key result crucially depend on the assumption that star formation is tightly correlated with H$_2$ abundance. The conclusions can therefore be subject to uncertainties associated with uncertainties of H$_2$ modelling. Note that one of the key ingredients of the H$_2$ formation model is H$_2$ formation rate on dust, for which we adopt the empirically constrained rate of \citet{wthk08} and do not assume any temperature dependence \citep[see Appendix in][for details]{ng:gk11}. It was argued, however, on theoretical grounds that such temperature dependence may exist \citep{burke_hollenbach83,cazaux_spaans09}. Both the degree of the temperature dependence and its effects on our conclusion are uncertain and such uncertainties of molecular hydrogen formation in environments drastically different from the Milky Way should be viewed as one of the caveats to our results and conclusions.  

The mass-metallicity relation for the model galaxies from our fiducial
simulation is shown in Figure \ref{fig:mz}. In concord with the
expectations presented in \S \ref{sec:theory}, the distribution of
simulated galaxies is clearly bimodal: galaxies with stellar masses
 $\gtrsim {\rm few\,}\times 10^8 \msol$ (residing in halos with total masses
larger than about $\sim 2\times 10^{10} \msol$) and metallicities
higher than $\approx 0.1Z_{\odot}$  comprise a distinctly different
population, with a much weaker scaling between stellar mass and
metallicity compared to galaxies with stellar masses $\lesssim 10^7
\msol$ and metallicities  $\lesssim 10^{-2}Z_{\odot}$. Galaxies
with intermediate values of stellar mass and metallicity are scarce,
but they appear to be following the well-defined extension of the $M_{\ast}-Z$ scaling of lower mass galaxies. 
Interestingly, the results do not change
significantly with decreasing redshift: the low-$M_{\ast}$,
low-$Z$ population does not become significantly depleted with
increasing cosmic time.  

Note that this bi-modality in the distribution of galaxy metallicities is a unique signature of 
H$_2$-based star formation, set by the sensitivity of H$_2$ abundance to the abundance of dust. 
Such bi-modality is not present in simulations
which use common star formation recipes based on the total gas density rather
than on the density of only molecular gas \cite[see, e.g., Fig. 6 in][]{tkg08}.
Dust is important both as a catalyst for H$_2$ formation and as a shield from dissociating UV radiation. As shown by \citet{ays11}, UV heating may also suppress formation of dense regions in the first place. Our results indicate that such suppression in high-redshift small halos is not complete. Overall, however, such heating contributes to increasing inefficiency of gas conversion into stars in such systems and thus is one of the reasons for the bi-modality that we observe.

Figure~\ref{Hist} demonstrates that bi-modality can also be apparent in the metallicity distribution of stars within individual galaxies. 
It shows the metallicity distribution of stellar particles in three galaxies of different stellar mass at
$z=3$. The galaxy in the left panel has a stellar mass of $3\times
10^{10} \msol$, and its distribution of stellar metallicities is bi-modal, 
although the fraction of stars in the low-metallicity peak is small. 
The distribution of metallicities in galaxy of intermediate stellar mass ($1.7 \times 10^8
\msol$), shown in the middle panel, exhibits distinct bimodality with the low-$Z$ peak containing a significant fraction of stars. 
Finally, the smallest galaxy with stellar mass of $1.5\times 10^{6} \msol$, shown in the right panel, has only a narrow low-$Z$ peak. 
This galaxy never evolves to undergo a transition to the fast star formation regime and thus never developes the high-metallicity peak present in larger mass galaxies (see discussion in the previous section).

\section{Comparison with observations}\label{comparison}

\begin{figure}
\plotone{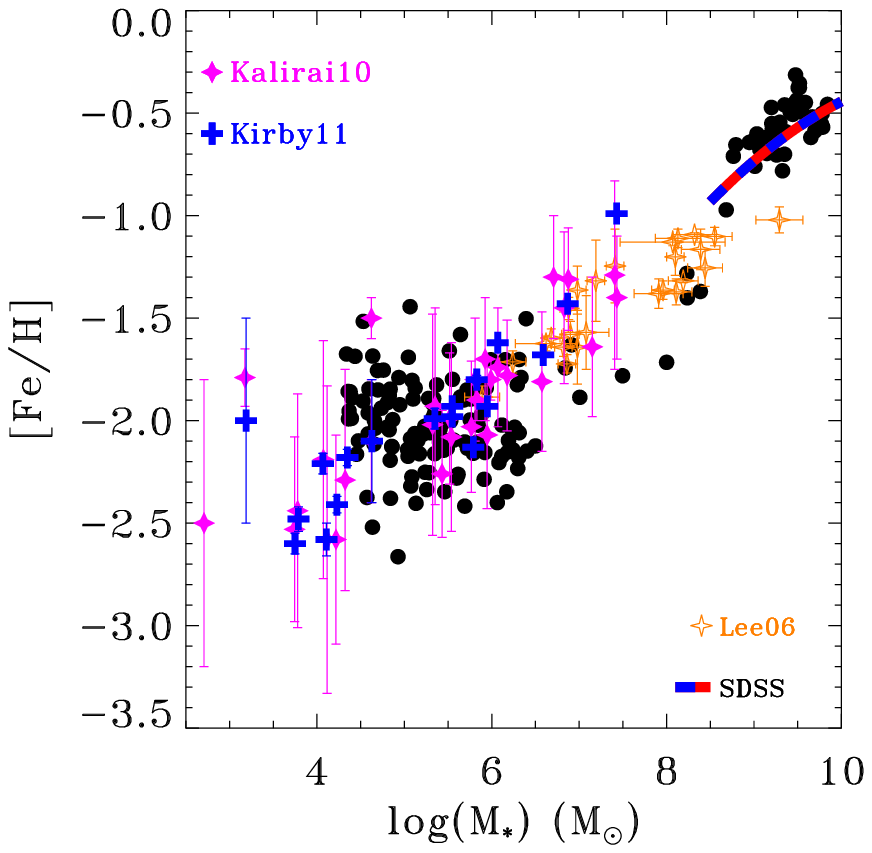}
\caption{\label{fig:mzobs} 
Stellar mass-metallicity relation for nearby dwarf galaxies. 
Blue crosses: iron metallicity observations in local dwarf galaxies from
\cite{dsh:klsc11}; magenta stars: \cite{kalirai10}; orange stars: dwarf galaxies from sample of \citet{lee06}. Average relation measured for the SDSS galaxies by \citet{Trem04} is shown by colored dashed line.
Simulation results from $z=3$, with added stochastic metallicity floor described by log-normal pdf with the median of $1\times 10^{-2}$ solar and scatter of 0.2~dex
are shown for comparison with the black solid dots. 
}
\end{figure}

Color points in Figure~\ref{fig:mzobs} show the $M_{\ast}-Z$ correlation for the Local Group dwarf galaxies \citep{kalirai10,dsh:klsc11} and other nearby dwarfs \citep{lee06} covering a mass range of $M_{\ast}\approx 10^3-10^8\rm\, \msol$ along with the mean relation for more massive galaxies \citep{Trem04}. The metallicities are measured for the Local Group dwarfs using stellar iron abundance, while \citet{lee06} and \citet{Trem04} measured gas oxygen abundances. We rescale the latter using the empirical prescription of \citet{simon_etal06} [eq.(12)]. After such rescaling both the Local Group dwarf spheroidals and dwarf irregular galaxies in the \citet{lee06} sample exhibit a fairly tight correlation that joins smoothly into the relation measured for the large sample of the SDSS galaxies by \citet{Trem04}. Note that $Z-M_{\ast}$ relation flattens substantially at the lowest stellar masses: at $M_{\ast}\lesssim 10^5\rm\,\msol$ metallicities no longer exhibit a tight correlation with stellar mass and are instead scattered in the range $Z\sim 10^{-2.5}-10^{-1.5}$. 

Even a cursory comparison of the predicted relation shown in Figure~\ref{fig:mz} with the correlation of real galaxies in  Figure~\ref{fig:mzobs} clearly shows that the two are inconsistent. The main difference is that simulated dwarf galaxies have drastically lower metallicities compared to the observed ones at a given stellar mass. This discrepancy can have different explanations and we discuss some of them below. Here we note that one possible explanation is that gas was universally pre-enriched to a metal floor of $Z_{\rm min}\sim 1\times 10^{-2}$. Such pre-enrichment could indeed explain the apparent lack of correlation between metallicity and stellar mass for the smallest dwarf galaxies. Given that different systems and regions in the early universe evolve somewhat differently, we can generally expect such pre-enrichment to be stochastic. Filled black circles in Figure~\ref{fig:mzobs} show simulated galaxies in which a stochastic metallicity floor was added, which we model as a log-normal with median metallicity of $0.01Z_{\odot}$ and scatter of $\sigma_{log_{10}Z}=0.2\dim{dex}$. Although the agreement of observations with such floor is not perfect, it is much improved compared to when the original simulation metallicities are used. Substantial pre-enrichment of gas of dwarf galaxy disks by preceding Pop III stars can thus help explain the discrepancy between predictions in Figure~\ref{fig:mz} and observations in Figure~\ref{fig:mzobs}. 

The distribution of metallicities of dwarf galaxies in Figure~\ref{fig:mzobs} also does not exhibit any clear bi-modality.  Observations do not seem to show a paucity of galaxies at stellar masses $M_{\ast}\sim 10^6-10^8\rm\,\msol$ and metallicities $Z\sim 0.01-0.1$. The interpretation of this discrepancy is less certain, given that the sample of dwarf galaxies shown in the figure is not volume limited. Nevertheless, the lack of bi-modality in the distribution of average metallicities of dwarf galaxies is consistent with the lack of corresponding bi-modality in the metallicity distribution of stars within galaxies.  These distributions have been measured for both the Milky
Way and for many of the dwarf spheroidal galaxies in the Local Group
\citep{dsh:bc05,dsh:hitb06,dsh:shtg10,dsh:bthi11,dsh:klsc11,dsh:cgah11}
with the statistical power that would be sufficient to uncover the
level of bi-modality seen in our simulations. However, none has been
detected so far.

We elaborate on the possible causes of such a stark disagreement between the
generic prediction of the molecular-gas-based star formation model and
observations in the next section.

\section{Discussion}\label{disc}

As we show in \S~\ref{sec:theory} and \ref{sims}, the assumption that star formation is tied to 
the molecular phase of the interstellar medium naturally leads to
bi-modality in the distribution of stellar masses and metallicities of dwarf galaxies. Such bi-modality 
is not observed in the available data for nearby dwarf galaxies and in metallicity distribution of stars 
within such galaxies.

Given the absence of bi-modality, which we argue is a generic feature
of H$_2$-based star formation, one can ask whether the assumption
about tight relation between star formation and molecular gas is
valid. Relaxing such assumption could perhaps modify the prediction
about pronounced bi-modality. Although in principle other models that
do not consider molecular gas as a primary indicator of star forming
gas are possible \citep[e.g.]{sfr:oml10}, we note that there is
currently good motivation for H$_2$ based star formation because star formation rate on $\sim 100$pc-$1$~kpc scales correlates strongly with H$_2$ mass in the corresponding region \citep{wong_blitz02,sfr:blwb08,sfr:gtgs10,sfr:blwb11}. Such star formation model also naturally explains the observed trend of
the Kennicutt-Schmidt relation 
with metallicity \citep{ng:gk10,sfr:rwc11,sfr:bljo11}. 

An alternative explanation for the observed lack of the bi-modality is
primordial pre-enrichment of the ISM and part of the IGM by the
first Population III stars and, perhaps, by rapid 
subsequent formation of Population II stars. Such
pre-enrichment is often dubbed ``the metallicity floor'', although the
observed distribution of metallicities in the lowest mass dwarf
spheroidals in the Local Group is better described by a lognormal
distribition with the median at $Z\approx 10^{-2}\Zsun$ and an
rms scatter of $0.2\dim{dex}$.
Indeed, theoretically it is more natural to expect some probability distribution density of initial metallicities, given that different regions collapsing into a galaxy may have had different evolutionary histories and enrichment levels \citep[see, e.g., Fig. 2 in][]{gals:wtna11}. We can thus expect metallicities to be somewhat different in different galaxies and in different regions within a given galaxy. 

We show that assuming a minimum metallicity stochastically varying according to the log-normal pdf with parameters given above significantly improves the agreement of simulated galaxies with observed low-mass dwarf galaxies. Although pre-enrichment to $Z\sim 10^{-2}Z_{\odot}$ may seem extreme, it is actually in line with current theoretical models. For example, a recent study by \citet{gals:wtna11} shows that Population III stars can enrich gas of the entire host halo to metallicity of $Z\sim 10^{-3}Z_{\odot}$. It is important, however, to distinguish mass-weighted average metallicity of all the pre-enriched gas and metallicity of the central regions that will be the sites of Population II star formation. The latter can be expected to be enriched to  metallicities considerably higher than average. Indeed, simulations of 
 \citet{gals:wtna11} show that although gas in dwarf halos at $z>10$ is enriched by Population III stars to average mass-weighted metallicities of $\sim 10^{-3}-10^{-2.5}Z_{\odot}$, the first Population II stars may form in the central regions of halos with metallicities of $Z_\ast\approx 10^{-2}\,Z_{\odot}$ or even higher (see their Figure~3).

The pre-enrichment hypothesis can explain the sharp decline in the
distribution of Milky Way halo stars below ${\rm [Fe/H]}\sim -3$, the
flattening of the mass-metallicity relation to a plateau at low
metallicities, as well as the absence of bi-modal behavior in
metallicities and other observed properties of the baryonic components of dwarf
galaxies. Indeed, if the gas in dense, star forming regions of early galaxies becomes pre-enriched by the 
first population of stars to a level of $\sim 10^{-2}Z_{\odot}$ the bi-modality of the galaxy metallicity 
distribution will become considerably less pronounced (see the blue line in the top panel of Fig.~\ref{fig:ansol}). 
 
A related observational fact is the lack of extremely low  metallicity ($Z<10^{-3}\Zsun$) Damped Ly-$\alpha$ (DLA) systems in current observational samples \citep[][Rafelski, M. et al. 2011, in preparation]{igm:pgwc03,igm:kphe10}. Observations are sufficiently sensitive to detect systems with metallicities as low as $\approx 10^{-4}\rm\,Z_\odot$, yet no systems with metallicity $\lesssim 3\times 10^{-2}\, Z_\odot$ have been yet found even in the highest redshift systems ($4\lesssim z\lesssim 5$). The lack of low-metallicity DLA systems also points towards significant early pre-enrichment of galactic disks. It remains to be seen whether simulations modeling pre-enrichment by Population III and II stars will be able to reproduce such fairly universal level of pre-enrichment in different stellar systems and gaseous disks (e.g., no place in the universe is clearly demonstrated to have metallicity much below $10^{-3}$ solar).

The hypothesis that most of the ultra-faint Local Group dwarfs ($M_* <
10^6\Msun$) are pre-enriched by the first generation of stars implies that essentially all heavy elements observed in systems with $M_\ast\lesssim 10^5\rm\,\msol$ have been produced by the Population III stars. This argument adds support to independent arguments for significant Population III pre-enrichment based on chemical evolution modelling of dwarf galaxies \citep{dsh:klsc11} and the Milky Way halo \citep{roll09}, and examination of abundance patterns in stars belonging to dwarfs \citep{simon10leo} and in damped Ly$\alpha$ systems \citep[][see also \citealp{kbbh11} for a recent review]{cooke11}. Measurements of the abundance patterns in spectra of stars within these galaxies can thus directly probe the nucleosynthetic yields of the very first supernovae. This opens an exciting avenue of using these smallest galaxies as unique window into the properties of first stars and conditions for star formation during the earliest stages of galaxy evolution.

\acknowledgements 

This work was supported in part by the DOE at Fermilab, by the NSF
grants AST-0507596 and AST-0708154, by the NASA grant NNX-09AJ54G, and
by the Kavli Institute for Cosmological Physics at the University of
Chicago through the NSF grant PHY-0551142 and an endowment from the
Kavli Foundation. The simulations used in this work have been
performed on the Joint Fermilab - KICP Supercomputing Cluster,
supported by grants from Fermilab, Kavli Institute for Cosmological
Physics, and the University of Chicago.
Part of this work was carried out at the Jet Propulsion Laboratory,
California Institute of Technology, under a contract with the National
Aeronautics and Space Administration.
This work made extensive use of the NASA
Astrophysics Data System and {\tt arXiv.org} preprint server.

\bibliographystyle{apj}
\bibliography{ms,ng-bibs/gals,ng-bibs/sfr,ng-bibs/self,ng-bibs/misc,ng-bibs/sims,ng-bibs/dsh,ng-bibs/ism,ng-bibs/igm,ng-bibs/cosmo}

\end{document}